\documentclass[conference,10pt]{IEEEtran}
\pdfoutput=1
\usepackage[boxruled]{algorithm2e}
\usepackage{authblk}
\usepackage{cite}
\usepackage{graphicx}
\usepackage{psfrag}
\usepackage{subfigure}
\usepackage{url}
\usepackage{amsmath}
\usepackage{array}
\usepackage{amssymb}
\usepackage{amsfonts}
\usepackage{graphicx}
\usepackage{epstopdf}
\usepackage{placeins}
\usepackage{algorithmic}

\newtheorem{lemma}{Lemma}

\newtheorem{definition}{Definition}
\newtheorem{theorem}{Theorem}

\newtheorem{example}{Example}

\usepackage{setspace}
\usepackage{amscd}
\usepackage{multirow}
\usepackage{mathrsfs}
\usepackage{epsfig}
\usepackage{color}
\usepackage{textcomp}
\usepackage{multirow}

\input{epsf.sty}

\title{Error Correcting Index Codes and Matroids}
\begin{document}
\author{Anoop Thomas and B. Sundar Rajan
}\affil{Dept. of ECE, IISc, Bangalore 560012, India, Email: $\lbrace$anoopt,bsrajan$\rbrace$@ece.iisc.ernet.in.}
\date{\today}
\maketitle
 \thispagestyle{empty}	
\begin{abstract}
The connection between index coding and matroid theory have been well studied in the recent past. El Rouayheb \textit{et al.} established a connection between multi linear representation of matroids and wireless index coding. Muralidharan and Rajan showed that a vector linear solution to an index coding problem exists if and only if there exists a representable discrete polymatroid satisfying certain conditions. Recently index coding with erroneous transmission was considered by Dau \textit{et al.}. Error correcting index codes in which all receivers are able to correct a fixed number of errors was studied. In this paper we consider a more general scenario in which each receiver is able to correct a desired number of errors, calling such index codes  \textit{differential error correcting index codes.} A link between differential error correcting index codes and certain matroids is established. We define  \textit{matroidal differential error correcting index codes} and we show that a scalar linear differential error correcting index code exists if and only if it is  matroidal differential error correcting index code associated with a representable matroid.
\end{abstract}

\section{Introduction}
\label{Sec:Introduction}

The index coding problem introduced by Birk and Kol \cite{ISCO} involves a source which generates a set of messages and a set of receivers which demand messages. Each receiver has prior knowledge of a portion of the message called side-information, which is known to the source. The source uses the knowledge of the side-information available at all the receivers to find a transmission scheme of minimum number of transmissions, which satisfies all the demands of the receivers. Bar-Yossef \textit{et al.} \cite{ICSI} studied the index coding problem and found that the length of the optimal linear index code is equal to the minrank of a related graph.
 
The connection between multi-linear representation of matroids and index coding was studied by El Rouayheb, Sprinston and Georghiades \cite{ICMT}. It was shown by Muralidharan and Rajan \cite{LICDPM} that a vector linear solution to an index coding problem exists if and only if there exists a representable discrete polymatroid satisfying certain conditions which are determined by the index coding problem.

The problem of index coding with erroneous transmissions was studied by Dau \textit{et al.} \cite{ECIC}. An index code capable of correcting at most $\delta$-errors at all its receivers is defined as a $\delta$-error correcting index code. The necessary and sufficient conditions for a scalar linear index code to have $\delta$-error correcting capability was found. Network linear network error-correcting codes were introduced earlier by Yeung and Cai \cite{NEC1},\cite{NEC2}. The link between network error correcting and matroid theory was established by Prasad and Rajan in \cite{MFNEC}. 

Index coding problems are modeled in the following way. There is a unique source $S$ having message $\mathbf{\mathit{x}}=(\mathit{x_{1}},\mathit{x_{2}}\ldots,\mathit{x_{n}}) \in \mathbb{F}_{q}^{n}$, where $\mathbb{F}_{q}$ is a finite field of $q$ elements. There are $m$ receivers $R_{1},R_{2},\ldots,R_{m}$. Let $\left\lceil n\right\rfloor$ denote the set $\left\{1,2,\ldots,n \right\}$. Each $R_{i}$ possesses a set of messages $\{x_{j}\}_{j\in \chi_{i}}$, where $\chi_{i}\subseteq \left\lceil n\right\rfloor$. Let $\chi=\lbrace \chi_{1},\chi_{2},\ldots, \chi_{m} \rbrace$ represents the set of side information possessed at all receivers. Each receiver $R_{i}$ demands a single message $x_{f(i)}$. The mapping $f:\left\lceil m\right\rfloor \rightarrow \left\lceil n\right\rfloor$ satisfies $f(i) \notin \chi_{i}$. So an index coding problem is described by the quadruple $(m,n,\chi,f)$. The objective is to find an encoding scheme referred to as an index code that satisfies all receivers and uses minimum number of transmissions.

\begin{definition}
\label{def:IC}
An \textit{index code} over $\mathbb{F}_{q}$ for an instance of the index coding problem described by $(m,n,\chi,f)$ is an encoding function $\mathfrak{C}:\mathbb{F}_{q}^{n} \rightarrow \mathbb{F}_{q}^N$ such that for each receiver $R_{i}$, $i \in \left\lceil m\right\rfloor$, there exists a decoding function $ \mathfrak{D}_{i}:\mathbb{F}_{q}^{N} \times \mathbb{F}_{q}^{|\chi_{i}|}\rightarrow \mathbb{F}_{q} $ satisfying $ \forall ~ x \in \mathbb{F}_{q}^{n} : \mathfrak{D}_{i}(\mathfrak{C}(x),x_{\chi_{i}})=x_{f(i)}$. The parameter $N$ is called the \textit{length} of the index code.

A \textit{linear index code} is an index code, for which the encoding function $\mathfrak{C}$ is a linear transformation over $\mathbb{F}_{q}$. Such a code can be described as $\forall ~x \in \mathbb{F}_{q}^{n} : \mathfrak{C}(x)=xL$ where $L$ is an $n\times N$ matrix over $\mathbb{F}_{q}$. The matrix $L$ is called the matrix corresponding to the linear index code $\mathfrak{C}$. The code $\mathfrak{C}$ is referred to as the linear index code based on $L$.
\end{definition}

Note that the encoding function $\mathfrak{C}:\mathbb{F}_{q}^{n} \rightarrow \mathbb{F}_{q}^N  $, can be viewed as $N$ encoding functions $c_{1},c_{2},\ldots,c_{N}$, where each $c_{i}$ is a function from $\mathbb{F}_{q}^{n} \rightarrow \mathbb{F}_{q}$. For the source message $x \in \mathbb{F}_{q}^n$, $c_{i}(x)$ gives the $i^{th}$ transmission of the index code. So an $N$ length index code $\mathfrak{C}$ can also be described by set of $N$ functions $\lbrace c_{1},c_{2},\ldots,c_{N} \rbrace$, denoted by $S(\mathfrak{C})$.

\begin{figure}
\centering{}
\includegraphics[scale=0.6]{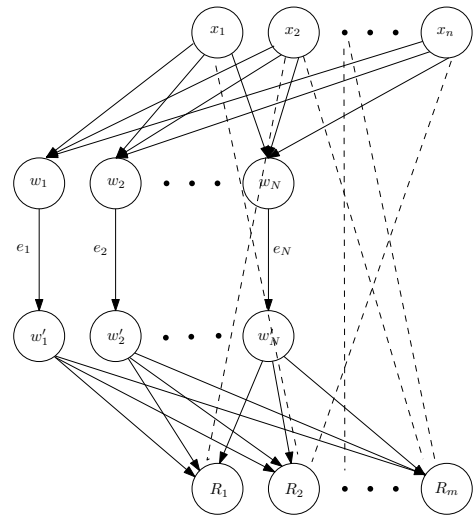}
\caption{Equivalent instance of network coding problem for an index coding problem $(m,n,\chi,f)$.}
\label{Fig:EquivalentNetworkCode}
\end{figure}

Rouayheb \textit{et al.} \cite{ICMT} proved that the index coding problem has a solution if and only if the equivalent network coding problem is solvable. The equivalent network coding problem for the index coding problem is as follows. Consider an instance of the index coding problem described by the quadruple $(m,n,\chi,f)$ with an index code $\mathfrak{C}$ of length $N$. The corresponding equivalent network coding problem for the index coding problem $(m,n,\chi,f)$ and index code $\mathfrak{C}$ is given in Fig. \ref{Fig:EquivalentNetworkCode}. The equivalent network coding problem has $n$ source nodes each corresponding to the source messages of the index coding problem. There are two levels of intermediate nodes denoted by $w_{i}$ and $w'_{i}$ for $i= 1$ to $N$. Each node $w_{i}$ has $n$ incoming edges, one from each of the source nodes. Each node $w'_{i}$ has only one input edge $e_{i}$, from $w_{i}$. The nodes $R_{1},R_{2},\ldots,R_{m}$ are the receiver nodes. Each receiver node $R_{i}$ has two types of input edges, $N$ input edges from each of the $w'_{i}$ node and there are $|\chi_{i}|$ input edges which correspond to the side information possessed by the receiver. The edges corresponding to side information are represented as dashed lines in the figure. The side information edges for a receiver $R_{i}$ are the edges $(x_{j},R_{i})$ for $j \in \chi_{i}$.  The information flowing through the edge $e_{i}$ of the equivalent network corresponds to the $i^{th}$ transmission of the index code $\mathfrak{C}$. The problem of index coding with error correction with receivers capable of correcting at most $\delta$ errors can be transformed to a network error correcting problem. Hence, a natural  question that arise is whether it is possible to link index codes and representable matroids by first transforming the index coding problem into equivalent network coding problem as in \cite{ICMT} and then link error correcting network code to matroids as in \cite{MFNEC}. If one takes this approach then, in the equivalent network, network code should be able to correct $m\delta$ errors as each receiver has $\delta$ error correcting capability. Also these errors are restricted to certain edges of the equivalent network. The errors can occur only in the edges between $w'_{i}$ and receiver nodes. Among the $N$ edges between $w'_{i}$ and $R_{j}$, at most $\delta$ can be in error. There are $m$ receivers and hence at most $m\delta$ errors.  Such case of restriction of errors to a subset of edges was not considered earlier. 
We also note that the size of the ground set of the matroid for which we have to find representation is smaller compared to the matroid obtained from the equivalent network coding problem. If we consider $m\delta$ errors without any restriction and equivalent network the size of the ground set of the  matroid obtained is $n+2|\mathcal{E}|$, where $\mathcal{E}$ is the edge set of the equivalent network. Number of edges in the equivalent network for an index coding problem $(m,n,\chi,f)$ is $|\mathcal{E}|= (n+m+1)N + \overset{m}{\underset{i=1}{\sum}} |\chi_{i}|$. The matroid which we obtain has a ground set of cardinality only $n+2N$ which is much smaller than $n+2|\mathcal{E}|$.

The contributions of this paper are as follows.
\begin{itemize}
\item We define differential error correcting index code in which the receivers  have different error correcting capability. The necessary and sufficient conditions for a matrix $L$ to correspond to a differential index code is found.
\item We establish a link between linear differential error correcting index codes and certain matroids. We define a matroidal differential error correcting index code and show that a linear differential error correcting index code exists if and only if it is matroidal differential error correcting index code associated with a representable matroid.
\item We consider two special cases, $\delta$-error correcting index codes and error correction at a subset of receiver nodes. The representable matroids associated with these two special cases are identified.
\end{itemize}

The organization of the paper is as follows. In Section \ref{Sec:ECIC} we review the definitions of error correcting index codes. We define differential error correcting index codes and also establishes a lemma which is used to prove our main result. In Section \ref{Sec:Matroids}, basic results of Matroids are reviewed. Finally in Section \ref{Sec:MECIC}, we define matroidal differential error correcting index code and establishes the link between matroids and differential error correcting index codes. We conclude and summarize all the results in Section \ref{Sec:Conclusion}.

\textit{Notations:}
The size of a set $S$ is denoted by $|S|$. Consider two sets $S_{1}$ and $S_{2}$. The set subtraction $S_{1} \setminus S_{2}$ is denoted by $S_{1}-S_{2}$. The rank of a matrix $A$ over $\mathbb{F}_{q}$ is denoted by $rank(A)$. For some positive integer $N$, identity matrix of size $N$ over $\mathbb{F}_{q}$ is denoted by $I_{N}$. The vector space spanned by columns of a matrix $A$ over $\mathbb{F}_{q}$ is denoted by $\langle A \rangle$. Consider a vector $x =(x_{1},x_{2},\ldots, x_{n}) \in \mathbb{F}_{q}^{n}$. The \textit{support} of vector $x$ is defined to be the set $supp(x)=\lbrace i \in \lceil n \rfloor : x_{i} \neq 0 \rbrace.$ The Hamming weight of a vector $x$, denoted by $wt(x)$, is defined to be the $|supp(x)|$. For some subset $B=\lbrace i_{1},i_{2},\ldots, i_{b} \rbrace$ of $\lceil n \rfloor $, where $i_{1} < i_{2} < \ldots < i_{b}$, let $x_{B}$ denote the vector $(x_{i_1},x_{i_2},\ldots,x_{i_b})$.  For some matrix $A$, $A^{(i)}$ denotes the $i^{th}$ column of $A$. For a set of column indices $\mathcal{I}$, $A^{\mathcal{I}}$ denotes the submatrix of $A$ with columns indexed by $\mathcal{I}$. Similarly $A_{(j)}$ denotes the $j^{th}$ row of $A$ and for a set of row indices $\mathcal{J}$, $A_{\mathcal{J}}$ denotes the submatrix of $A$ with rows indexed by $\mathcal{J}$.

\section{Differential Error Correcting Index Codes and a useful Lemma}
\label{Sec:ECIC}

In this section, we review error correcting index codes, define  differential error correcting index code and establish a lemma which we use in Section \ref{Sec:MECIC} to establish our main result.


Error correcting index codes consider the scenario in which the symbols received by receiver $R_{i}$ may be subject to errors. The source $S$ broadcasts a vector $\mathfrak{C}(x) \in \mathbb{F}_{q}^{N}$. The error affecting receiver $R_{i}$ is considered as an additive error represented by $\epsilon_{i} \in \mathbb{F}_{q}^{N}$. Then, $R_{i}$ actually receives the vector $y_{i}=\mathfrak{C}(x)+\epsilon_{i} \in \mathbb{F}_{q}^{N}$. An error correcting index code should be able to satisfy the demands of receivers in the presence of these additive errors.

\begin{definition}
\label{def:ECIC}
Consider an instance of the index coding problem described by $(m,n,\chi,f)$. A \textit{$\delta$-error correcting index code} ($\delta$-ECIC) over $\mathbb{F}_{q}$ for this instance is an encoding function $ \mathfrak{C}:\mathbb{F}_{q}^{n} \rightarrow \mathbb{F}_{q}^N $ such that for receiver $R_{i}$, $i \in \left\lceil m\right\rfloor$, there exists a decoding function $\mathfrak{D}_{i}:\mathbb{F}_{q}^{N} \times \mathbb{F}_{q}^{|\chi_{i}|}\rightarrow \mathbb{F}_{q} $ satisfying $ \forall ~ x \in \mathbb{F}_{q}^{n}, \forall ~ \epsilon_{i} \in \mathbb{F}_{q}^{N}, wt(\epsilon_{i})\leq \delta : \mathfrak{D}_{i}(\mathfrak{C}(x)+ \epsilon_{i},x_{\chi_{i}})=x_{f(i)}$.

\end{definition}

If the encoding function $\mathfrak{C}$ is linear then it is a \textit{linear error correcting index code}. As for the linear index code, linear error correcting index code can also be described by a matrix $L$. 

Dau \textit{et al.} in \cite{ECIC} identify a necessary and sufficient condition which a matrix $L$ has to satisfy to correspond to a $\delta$-error correcting index code. Consider the set of vectors  \[ \mathcal{I}(q,\mathcal{H})\triangleq \lbrace z \in \mathbb{F}_{q}^{n} : \exists ~ i \in \lceil m \rfloor ~ | ~ z_{\chi_{i}}=0 \text{ and } z_{f(i)}\neq 0\rbrace.\] Here $\mathcal{H}$ is a directed hypergraph \cite{BCSI} used to describe the index coding problem $(m,n,\chi,f)$.  The necessary and sufficient conditions for a matrix $L$ to correspond to a $\delta$-error correcting index code is as follows. The matrix $L$ corresponds to a $(\delta,\mathcal{H})$-ECIC over $\mathbb{F}_{q}$ if and only if 
\begin{equation}
\label{eq:wtcriterion}
wt(zL)\geq 2\delta+1, \forall ~ z \in \mathcal{I}(q,\mathcal{H}).
\end{equation}


The proof of \eqref{eq:wtcriterion} follows from the fact that if \eqref{eq:wtcriterion} is not satisfied  then one can demonstrate the existence of a pair of information vectors $x$ and $x'$ with $x_{f(i)}\neq x'_{f(i)}, x_{\chi_{i}}=x'_{\chi_{i}}$ and a corresponding pair of error vectors $\epsilon$ and $\epsilon'$ with  $wt(\epsilon)\leq \delta$ and $wt(\epsilon')\leq \delta$ such that the corresponding received vectors at receiver $R_{i}$ are equal. So the receiver $R_{i}$ will not be able to distinguish between $x_{f(i)}$ and $x'_{f(i)}$. Therefore, $L$ corresponds to a $\delta$-error correcting index code if and only if the following condition holds. For all $i \in \lceil m \rfloor$ and for all $z \in \mathbb{F}_{q}^{n}$ such that $z_{\chi_{i}}=0$ and $z_{f(i)} \neq 0$,
$zL + e \neq 0, \forall ~ e \in \mathbb{F}_{q}^{N}, wt(e) \leq 2\delta$. 
This equation can be rewritten in matrix form in the following way. For each receiver $R_{i}$
\begin{flalign}
\begin{split}
\label{eq:ECMatrixRepresentation}
(z\quad e)\left(\begin{array}{c}
L\\
I_{N}
\end{array}\right)\neq 0, \forall ~ z \in \mathbb{F}_{q}^{n} : z_{\chi_{i}}=0,z_{f(i)} \neq 0, \\ \forall ~ e \in \mathbb{F}_{q}^{N} : wt(e)\leq 2\delta.
\end{split}
\end{flalign} 

The matrix $L$ corresponds to an error correcting index code if and only if  \eqref{eq:ECMatrixRepresentation} is satisfied at all receivers  $R_{i},$  $i \in \lceil m \rfloor$.

We now define a \textit{differential $\delta_{R}$ error correcting index code}. A differential error correcting index coding problem considers the scenario in which the error correcting capability varies from receiver to receiver. 
\begin{definition}
\label{def:DECIC}
Consider an instance of the index coding problem described by $(m,n,\chi,f)$. Let $\delta_{R}=\lbrace \delta_{1},\delta_{2},\ldots,\delta_{N} \rbrace $, where $\delta_{i}$ is the maximum number of errors receiver $R_{i}$ wants to correct. A \textit{differential $\delta_{R}$ error correcting index code} over $\mathbb{F}_{q}$ for this instance is an encoding function $ \mathfrak{C}:\mathbb{F}_{q}^{n} \rightarrow \mathbb{F}_{q}^N $ such that for receiver $R_{i}$, $i \in \left\lceil m\right\rfloor$, there exists a decoding function $ \mathfrak{D}_{i}:\mathbb{F}_{q}^{N} \times \mathbb{F}_{q}^{|\chi_{i}|}\rightarrow \mathbb{F}_{q} $ satisfying $ \forall ~ x \in \mathbb{F}_{q}^{n}, \forall ~ \epsilon_{i} \in \mathbb{F}_{q}^{N}, wt(\epsilon_{i})\leq \delta_{i} : \mathfrak{D}_{i}(\mathfrak{C}(x)+ \epsilon_{i},x_{\chi_{i}})=x_{f(i)}$.
\end{definition}

Equation \eqref{eq:ECMatrixRepresentation} needs to be modified for the differential $\delta_{R}$ error correcting index code. A matrix $L$ corresponds to a differential error correcting index code if and only if the following condition holds. For each receiver $R_{i}, i \in \lceil m \rfloor$,
\begin{flalign}
\begin{split}
\label{eq:DECMatrixRepresentation}
(z\quad e)\left(\begin{array}{c}
L\\
I_{N}
\end{array}\right)\neq 0, \forall ~ z \in \mathbb{F}_{q}^{n} : z_{\chi_{i}}=0,z_{f(i)} \neq 0, \\ \forall ~ e \in \mathbb{F}_{q}^{N} : wt(e)\leq 2\delta_{i}.
\end{split}
\end{flalign} 

The \textit{error pattern} corresponding to an error vector $e$ is defined as its support set $supp(e)$. Let $\mathbb{I}_{supp(e)}$ denote the submatrix of $I_{N}$ consisting of those rows of $I_{N}$ indexed by $supp(e)$. For a receiver $R_{i}$, the error correcting condition \eqref{eq:DECMatrixRepresentation} can be rewritten as 
\begin{equation}
\begin{split}
\label{eq:EC2}	 	
(z\quad \overline{e})\left(\begin{array}{c}
L\\
\mathbb{I}_{supp(e)} 
\end{array}\right)\neq 0, \forall ~ z \in \mathbb{F}_{q}^{n}: z_{\chi_{i}} = 0,z_{f(i)} \neq 0, \\ \forall ~ \overline{e} \in \mathbb{F}_{q}^{2\delta_{i}}, \forall ~ supp(e) \in \lbrace \mathcal{F}\subseteq \lceil N \rfloor : |\mathcal{F}|=2\delta_{i} \rbrace.
\end{split}
\end{equation}
Let $\overline{\chi_{i}}$ denote the set $\lceil n \rfloor - \chi_{i}$. Since at a particular receiver we consider only those $z \in \mathbb{F}_{q}^{n}$ for which $z_{\chi_{i}}=0$  condition (\ref{eq:EC2}) can be rewritten as 
\begin{flalign}
\begin{split}
\label{eq:EC3}
(z_{\overline{\chi_{i}}}\quad \overline{e}) & \left(\begin{array}{c}
L_{\overline{\chi_{i}}}\\
\mathbb{I}_{supp(e)}
\end{array}\right)\neq 0, \forall ~ z_{\overline{\chi_{i}}} \in \mathbb{F}_{q}^{|\overline{\chi_{i}}|} : z_{f(i)} \neq 0, \\ & \forall ~ \overline{e} \in \mathbb{F}_{q}^{2\delta_{i}}, \forall ~ supp(e) \in \lbrace \mathcal{F}\subseteq \lceil N \rfloor : |\mathcal{F}|=2\delta_{i} \rbrace.
\end{split}
\end{flalign}
We now present a lemma which will be used in Section \ref{Sec:MECIC} to prove the main result of this paper.

\begin{lemma}
\label{lem:errorCorrection}
Let $\mathcal{I}_{f(i)}$ denote $(|\overline{\chi_{i}}|+2\delta_{i})$ length vector with a $1$ in one of the first $|\overline{\chi_{i}}|$ position corresponding to the demand $f(i)$ of the receiver $R_{i}$, and with all other elements $0$. For some $ supp(e) \in \lbrace \mathcal{F}\subseteq \lceil N \rfloor : |\mathcal{F}|=2\delta_{i} \rbrace$ the condition 
\begin{equation}
\begin{split}
(z\quad \overline{e})\left(\begin{array}{c}
L\\
\mathbb{I}_{supp(e)}
\end{array}\right)\neq 0, \forall ~ z \in \mathbb{F}_{q}^{n}: z_{\chi_{i}} = 0,z_{f(i)} \neq 0,\\ \forall ~ \overline{e} \in \mathbb{F}_{q}^{2\delta_{i}}
\end{split}
\end{equation}
holds if and only if the following condition holds 
\begin{equation}
\label{eq:EC4}
\mathcal{I}_{f(i)}\subseteq \left\langle \left(\begin{array}{c}
L_{\overline{\chi_{i}}}\\
\mathbb{I}_{supp(e)}
\end{array}\right)\right\rangle. 
\end{equation}

\begin{IEEEproof}
Refer Appendix \ref{App:LemmaProof}.
\end{IEEEproof}

\end{lemma}

Lemma \ref{lem:errorCorrection} gives an equivalent condition for equation \eqref{eq:EC2}. Therefore a given index code is differential $\delta_{R}$ error correcting if and only if \eqref{eq:EC4} holds for all $ supp(e) \in \lbrace \mathcal{F}\subseteq \lceil N \rfloor : |\mathcal{F}|=2\delta_{i} \rbrace$ and at all receivers $R_{i}, i \in \lceil m \rfloor$.

\section{Matroids}
\label{Sec:Matroids}
In this section we list few basic definitions and results from matroid theory. These results and definitions are taken from \cite{Oxley}.

\begin{definition}
\label{def:Matroid}
Let $E$ be a finite set. A matroid $\mathcal{M}$ on $E$ is an ordered pair $(E,\mathcal{I})$, where the set $\mathcal{I}$ is a collection of subsets of $E$ satisfying the following three conditions
\begin{list}{}
\item (I1) $\phi \in \mathcal{I}$
\item (I2) If $X \in \mathcal{I}$ and $X' \subseteq X$, then $X' \in \mathcal{I}$.
\item (I3) If $X_{1}$ and $X_{2}$ are in $\mathcal{I}$ and $|X_{1}| < |X_{2}|$, then there is an element $e \in X_{2} - X_{1}$ such that $X_{1} \cup e \in \mathcal{I}$.
\end{list}
\end{definition}

The set $E$ is called the \textit{ground set} of the matroid and is also referred to as $E(\mathcal{M})$. The members of set $\mathcal{I}$ are called the independent sets of $\mathcal{M}$. Independent sets are also denoted by $\mathcal{I}(\mathcal{M})$. A maximal independent subset of $E$ is called a \textit{basis} of $\mathcal{M}$ and the set of all bases of $\mathcal{M}$ is denoted by $\mathcal{B}(\mathcal{M})$. With $\mathcal{M}$, a function called the \textit{rank} function is associated, whose domain is the power set of $E$ and codomain is the set of non-negative integers. The rank of any $X \subseteq E$ in $\mathcal{M}$, denoted by $r_{\mathcal{M}}(X)$ is defined as the maximum cardinality of a subset $X$ that is a member of $\mathcal{I}(\mathcal{M})$.  The rank of matroid is the rank of its ground set.

\begin{definition}
\label{def:isomorphismMatroids}
Two matroids $\mathcal{M}_{1}$ and $\mathcal{M}_{2}$ are isomorphic, denoted as $\mathcal{M}_{1} \cong \mathcal{M}_{2}$, if there is a bijection $\psi$ from $E(\mathcal{M}_{1})$ to $E(\mathcal{M}_{2})$ such that, for all $X \subseteq E(\mathcal{M}_{1}),\psi(X)$ is independent in $\mathcal{M}_{2}$ if and only if $X$ is independent in $\mathcal{M}_{1}$.
\end{definition}

\begin{definition}
\label{def:vectormatroid}
The \textit{vector matroid} associated with a matrix $A$ over some field $\mathbb{F}$, denoted by $\mathcal{M}[A]$, is defined as the ordered pair $(E,\mathcal{I})$ where $E$ consists of the set of column labels of $A$, and $\mathcal{I}$ consists of all the subsets of $E$ which index columns that are linearly independent over $\mathbb{F}$. An arbitrary matroid $\mathcal{M}$ is said to be $\mathbb{F}$-\textit{representable} if it is isomorphic to a vector matroid associated with some matrix $A$ over some field $\mathbb{F}$. The matrix $A$ is then said to be a \textit{representation} of $\mathcal{M}$. A matroid which is not representable over any finite field is called a \textit{non-representable} matroid.
\end{definition}

Consider a matroid $\mathcal{M}$ with matrix $A$ as its representation. Each element in ground set of $\mathcal{M}$ corresponds to a column in $A$. For a subset $S$ of ground set $E(\mathcal{M})$, $A^{S}$ denotes the submatrix of $A$ with columns corresponding to the elements of ground set in $S$.  

\begin{lemma}
\label{lem:MatroidRowColumnOps}
Let $\mathcal{M}$ be a representable vector matroid with $A$ as its representation matrix over field $\mathbb{F}$. The matroid $\mathcal{M}$ remains unchanged if any of the following operations are performed on $A$
\begin{itemize}
\item Interchange two rows.
\item Multiply a row by a non-zero member of $\mathbb{F}$.
\item Replace a row by the sum of that row and another.
\item Adjoin or delete a zero row.
\item Multiply a column by a non-zero member of $\mathbb{F}$.

\end{itemize}
\end{lemma}




\begin{definition}
\label{def:Matroidcontraction}
Let $\mathcal{M}=(E,\mathcal{I})$ be a matroid and $T \subseteq E$. Consider the set $\mathcal{I}'=\lbrace I \subseteq E-T : I \cup B_{T} \in \mathcal{I} \rbrace$, where $B_{T} \subseteq T$ is a maximal independent subset within $T$. The contraction of $T$ from $\mathcal{M}$, denoted as $\mathcal{M}/ T$, is the matroid $(E - T,\mathcal{I}')$.
\end{definition}

The contraction of a $\mathbb{F}$-representable matroid is also $\mathbb{F}$-representable. Let $\mathcal{M}[A]$ be the vector matroid associated with a matrix $A$ over $\mathbb{F}$. Let $e$ be the index of a non-zero column of $A$. Suppose using the elementary row operations, we transform $A$ to obtain a matrix $A'$ which has a single non-zero entry in column $e$. Let $A''$ denote the matrix which is obtained by deleting the row and column containing the only non-zero entry of column $e$. Then $
\mathcal{M}[A]/\lbrace e \rbrace = \mathcal{M}[A'']. 
$

\section{Matroidal Error Correcting Index Codes}
In this section we first define matroidal differential error correcting index codes and then prove the main result of the paper.
\label{Sec:MECIC}

\begin{definition}
\label{def:MECIC}
Consider an index coding problem $(m,n,\chi,f)$ with an index code $\lbrace c_{1},c_{2},\ldots,c_{N} \rbrace$ of length $N$. Let $\delta_{R}=\lbrace \delta_{1},\delta_{2},\ldots,\delta_{N} \rbrace $, where $\delta_{i}$ is the maximum number of errors receiver $R_{i}$ wants to correct. Let $\mathcal{M}=(E,\mathcal{I})$ be a matroid  defined over a ground set $E$ with $n + 2N$ elements and with $r(\mathcal{M})= n + N$. The index code $\lbrace c_{1},c_{2},\ldots,c_{N} \rbrace$ is said to be \textit{matroidal differential $\delta_{R}$ error correcting index code} associated with $\mathcal{M}$, if there exists a function $g: \lceil n \rfloor \cup \lbrace c_{1},c_{2},\ldots,c_{N} \rbrace \rightarrow E(\mathcal{M})$ such that the following conditions are satisfied.
\begin{list}{}{}
\item (A) $g$ is one-one on $\lceil n \rfloor$, and $g(\lceil n \rfloor) \in \mathcal{I}(\mathcal{M})$.
\item (B) For at least one basis $B$ of $\mathcal{M}$ obtained by extending $g(\lceil n \rfloor)$, i.e. $B - g(\lceil n \rfloor)=\lbrace b_{n+1},b_{n+2},\ldots,b_{n+N} \rbrace$, the following conditions should hold.
\begin{list}{•}{•}
\item (B1) $g(c_{i}) \notin cl_{\mathcal{M}}(B-g(\lceil n \rfloor)), \forall ~i=1,2,\ldots,N.$
\item \begin{flalign*} \text{(B2) } r_{\mathcal{M}}(g(\lceil n \rfloor) \cup b_{n+i} \cup g(c_{i})) & = r_{\mathcal{M}}(g(\lceil n \rfloor) \cup b_{n+i}) \\ & =r_{\mathcal{M}} (g(\lceil n \rfloor) \cup g(c_{i})).\end{flalign*}
\end{list}
\item (C) For each receiver $R_{i}$ and for each error pattern $\mathcal{F}=\lbrace e_{i_{1}},e_{i_{2}},\ldots,e_{i_{2\delta_{i}}}\rbrace$, let \[
B_{\overline{\mathcal{F}},i}=B - g(\overline{\chi_{i}}) - \lbrace b_{n+i_{1}},b_{n+i_{2}},\ldots,b_{n+i_{2\delta_{i}}} \rbrace. \] Let $\mathcal{M}_{\mathcal{F},i}$ be the $|\overline{\chi_{i}}| + N + 2\delta_{i}$ element matroid $\mathcal{M}/B_{\overline{\mathcal{F}},i}$. Then at every receiver $R_{i}$ and for each valid error pattern $ \mathcal{F}$ we must have 
\begin{flalign*}
r_{\mathcal{M}_{\mathcal{F},i}}(g(\lbrace c_{1},c_{2},\ldots,c_{N} & \rbrace) \cup g(f(i)))\\ & =r_{\mathcal{M}_{\mathcal{F},i}}(g(\lbrace c_{1},c_{2},\ldots,c_{N} \rbrace)).
\end{flalign*}
\end{list} 
\end{definition}

Definition \ref{def:MECIC} can be viewed as a matroidal abstraction of differential $\delta_{R}$ error correcting index codes. Condition (A) ensures that the vectors representing the messages are linearly independent. Condition (B) is equivalent to the condition that scalar linear index code should be a non-zero linear combination of messages added with a linear combination of errors. Condition (C) ensures that the receivers can decode their demands in the presence of errors. Condition (C) requires that $g(\lbrace c_{1},c_{2},\ldots,c_{N} \rbrace)$ and $g(f(i))$ is present in $E(\mathcal{M}_{\mathcal{F},i})$. However this is ensured because condition (B) ensures $g(\lbrace c_{1},c_{2},\ldots,c_{N} \rbrace)$ does not belong to $B$ and hence it is not contracted out. Also since $g(f(i)) \subseteq g(\overline{\chi_{i}})$, it is also not contracted out of the matroid $\mathcal{M}$. We now present the main result of this paper which relates scalar linear differential error correcting index codes to representable matroids.

\begin{theorem}
\label{thm:MatroidalErrorCorrecting}
Let $(m,n,\chi,f)$ be an index coding problem. A scalar linear differential $\delta_{R}$ error correcting index code over $\mathbb{F}_{q}$ exists if and only if the index code is matroidal differential $\delta_{R}$ error correcting associated with a $\mathbb{F}_{q}$ representable matroid.

\begin{IEEEproof}
Refer Appendix \ref{App:TheoremProof}.
\end{IEEEproof}
\end{theorem}

Theorem \ref{thm:MatroidalErrorCorrecting} establishes a link between scalar linear differential $\delta_{R}$ error correcting index codes and a representable matroid satisfying certain properties. In the examples below, we consider differential error correcting  index coding problems with a scalar linear solution and show the representable matroid associated with it. 

\begin{example}
\label{eg:differentialerrorcorrecting1}
Let $q=2,m=n=3$ and $f(i)=i, \forall ~i \in \lceil m \rfloor$. Let $\chi_{1}=\lbrace 2 \rbrace,\chi_{2}=\lbrace 1,3 \rbrace,\chi_{3}=\lbrace 2,1 \rbrace$. Let $\delta_{1}=2$ and $\delta_{2}=\delta_{3}=1$. Consider the index code $\mathfrak{C}$ of length $N=7$ described by the matrix described by the matrix \[
L=\left[\begin{array}{ccccccccc}
1 & 1 & 1 & 1 & 0 & 1 & 0 \\
0 & 0 & 1 & 0 & 1 & 1 & 0 \\
0 & 1 & 0 & 1 & 1 & 0 & 1  
\end{array}\right]
.\]We have $\mathfrak{C}(x)=\lbrace x_{1},x_{1}+x_{3},x_{1}+x_{2},x_{1}+x_{3},x_{2}+x_{3},x_{1}+x_{2},x_{3}\rbrace$. The index code $\mathfrak{C}$ can also be described by seven encoding functions $S(\mathfrak{C})=\lbrace c_{1},c_{2},\ldots,c_{7} \rbrace$. Consider the representable matroid $\mathcal{M}$ associated with $10 \times 17$ matrix $
A=\left[
\begin{array}{cccccccc}
& L\\
I_{10}&\\
& I_{7} \\
\end{array}
\right]
$. The function $g: \lceil 3 \rfloor \cup \lbrace c_{1},c_{2},\ldots,c_{N} \rbrace \rightarrow E(\mathcal{M})$, is defined as follows,  \[ g(i)=i, \forall ~ i \in \lceil 3 \rfloor \]
\[ g(c_{i})=10+i, \forall ~ i \in \lceil 7 \rfloor. \] The matroid $\mathcal{M}$ and function $g$ satisfies Definition \ref{def:MECIC}. The first three columns corresponding to messages are linearly independent. Thus Condition (A) is satisfied. Consider the basis $B= \lbrace b_{1},b_{2},\ldots, b_{10}\rbrace = \lceil 10 \rfloor$. The set $B-g(\lceil 3 \rfloor)=\lbrace4,5,\ldots,10\rbrace$. The vectors corresponding to $g(c_{i})$ does not lie in the column space of the matrix $A^{(B - g(\lceil 3 \rfloor))}$. Thus Condition (B1) is satisfied. Condition (B2) is actually seven conditions corresponding to each $c_{i}, i \in \lceil 7 \rfloor$. For $c_{1}$, we have, \[
A^{(g(\lceil 3 \rfloor) \cup b_{4} \cup g(c_{1}))}= \left[\begin{array}{ccccccccc}
1 & 0 & 0 & 0 & 1 \\
0 & 1 & 0 & 0 & 0 \\
0 & 0 & 1 & 0 & 0 \\
0 & 0 & 0 & 1 & 1 \\
& & \mathcal{O} & & \\
\end{array}\right],
\] where $\mathcal{O}$ is a $6 \times 6$ all zero matrix. It is clear that the fifth column of $A^{g(\lceil 3 \rfloor) \cup b_{4} \cup g(c_{1}))}$ is a linear combination of the first four columns and also the fourth column is a linear combination of the remaining four columns. Similarly condition (B2) can be verified for all $c_{i}, i \in \lceil 7 \rfloor$. 

Condition (C) needs to be verified at all receivers. Let us consider receiver $R_{1}$ with $\delta_{1}=2$. Consider an error pattern $\mathcal{F}_{1}=\lbrace e_{1},e_{2},e_{3},e_{4}\rbrace$. We have $B_{\overline{\mathcal{F}_{1}},1}=B - g(\overline{\chi_{1}}) - \lbrace 4,5,6,7 \rbrace=\lbrace 2,8,9,10 \rbrace$. The matroid $\mathcal{M}_{\mathcal{F}_{1,1},1}=\mathcal{M}/B_{\overline{\mathcal{F}_{1,1}},1}$, is a vector matroid. Let $M_{i,j}$ be the matrix used for representation of the matroid $\mathcal{M}_{\mathcal{F}_{i,j},i} $. The vector representation for this matroid is given by the matrix 
\[
M_{1,1}=\left[
\begin{array}{cccccccc}
& 1 & 1 & 1 & 1 & 0 & 1 & 0\\
& 0 & 1 & 0 & 1 & 1 & 0 & 1\\
& 1 & 0 & 0 & 0 & 0 & 0 & 0\\
I_{6} & & & & & & & \\
& 0 & 1 & 0 & 0 & 0 & 0 & 0\\
& 0 & 0 & 1 & 0 & 0 & 0 & 0\\
& 0 & 0 & 0 & 1 & 0 & 0 & 0\\
\end{array}
\right]
.\]
In order to satisfy Condition (C), we want the columns of $M_{1,1}$ corresponding to $g(f(1))$ to be in the linear span of columns of $M_{1,1}$ corresponding to $g(S(\mathfrak{C}))$. The corresponding columns are given below: \[
M_{1,1}^{g(f(1))}=\left[\begin{array}{c}
1 \\
0 \\
0 \\
0 \\
0 \\
0
\end{array}
\right],M_{1,1}^{g(S(\mathfrak{C}))}=\left[\begin{array}{cccccccc}
1 & 1 & 1 & 1 & 0 & 1 & 0\\
0 & 1 & 0 & 1 & 1 & 0 & 1\\
1 & 0 & 0 & 0 & 0 & 0 & 0\\
0 & 1 & 0 & 0 & 0 & 0 & 0\\
0 & 0 & 1 & 0 & 0 & 0 & 0\\
0 & 0 & 0 & 1 & 0 & 0 & 0\\
\end{array}\right]
.\]We can observe that columns corresponding to $g(f(1))$ lies in the linear span of columns corresponding to $g(S(\mathfrak{C}))$. Condition (C) is verified for receiver $R_{1}$ and error pattern $\lbrace e_1,e_2,e_3,e_4 \rbrace$. Similarly it can be verified for all error patterns and for all receivers. Thus the matroid $\mathcal{M}$ with representation $A$ and function $g$ satisfies Definition \ref{def:MECIC}. So the index code $\mathfrak{C}$ is a matroidal differential $\delta_{R}$ error correcting index code. It can be verified that the index code $\mathfrak{C}$ is differential $\delta_{R}$ error correcting.
\end{example}

\begin{example}
\label{eg:example3}
Let $q=2,m=n=5$ and $f(i)=i, \forall ~i \in \lceil m \rfloor$. Let $\chi_{1}=\lbrace 2,5 \rbrace,\chi_{2}=\lbrace 1,3 \rbrace,\chi_{3}=\lbrace 2,4 \rbrace,\chi_{4}=\lbrace 3,5 \rbrace$ and $\chi_{5}=\lbrace 1,4 \rbrace$. Let $\delta_{1}=\delta_{4}=\delta_{5}=1$ and $\delta_{2}=\delta_{3}=2$. Consider the index code $\mathfrak{C}$ of length $N=8$, described by the matrix \[
L=\left[\begin{array}{ccccccccc}
1 & 1 & 1 & 1 & 0 & 0 & 0 & 0 \\
0 & 1 & 1 & 1 & 0 & 1 & 1 & 0 \\
1 & 1 & 0 & 0 & 1 & 1 & 1 & 0 \\
0 & 1 & 0 & 0 & 1 & 0 & 1 & 1 \\
1 & 0 & 0 & 1 & 0 & 0 & 1 & 1  
\end{array}\right]
.\] The index code $\mathfrak{C}$ can be viewed as eight encoding functions $S(\mathfrak{C})=\lbrace c_{1},c_{2},\ldots,c_{8} \rbrace$. Consider a representable matroid $\mathcal{M}$ associated with $13 \times 21$ matrix \[
A=\left[
\begin{array}{cccccccc}
& L\\
I_{13}&\\
& I_{8} \\
\end{array}
\right]
.\] The function $g: \lceil 5 \rfloor \cup S(\mathfrak{C}) \rightarrow E(\mathcal{M})$, is defined as follows,  \[ g(i)=i, \forall ~ i \in \lceil 5 \rfloor \]
\[ g(c_{i})=13+i, \forall ~ i \in \lceil 8 \rfloor. \] One can verify that the matroid $\mathcal{M}$ and function $g$ satisfies Definition \ref{def:MECIC}. Note that the length of this index code is eight. In \cite{ECIC}, for this index coding problem with all receivers requiring double error correction, the optimal length of index code was found to be 9. However since here few receivers require only single error correcting capability we are able to find the code of length eight. The restrictions on the matroid for which we need to find the representation are less. This enables us to find representations of smaller size. The size of the ground set of matroid is reduced to 21. For a double error correcting index code the size of the ground set is 23. 
\end{example}

\subsection{$\delta$-Error Correcting Index Codes}
Here we consider $\delta$-error correcting index codes in which all the receivers have the ability to correct $\delta$ number of errors. This is a special case of differential $\delta_{R}$ error correcting index code in which $\delta_{i}=\delta, \forall i \in \lceil m \rfloor$. We consider a single error correcting index coding problem and show the matroids associated with it in Example \ref{eg:example1} in Appendix \ref{App:ExampleErrorCorrectingIndexCodes}. 

\subsection{Error Correction at only a subset of Receivers}

Now we consider another special case of error correcting codes with error correcting capability at a subset of receivers. Consider a subset $S$ of $\lceil m \rfloor$. Each receiver $R_{i}, i \in S$ should be able to correct $\delta_{i}$ errors. We can obtain error correcting only at particular subset of receivers, from a differential error correcting index code by setting $\delta_{i}=0, \forall i \notin S$. Note that a $\delta$-error correcting index code corrects $\delta$ errors at all receivers. So it obviously corrects errors at a subset of receivers. However since the conditions which the matroid to satisfy are less we will be able to find other representations for the matroid. We illustrate this in Example \ref{eg:example2} in Appendix \ref{App:ExampleErrorCorrectionSubsetofReceivers}.

\section{Conclusion}
\label{Sec:Conclusion}
In this paper we have defined differential error correcting index codes which is a generalization of error correcting index codes. We established the connections between differential error correcting index codes and matroids. It was shown that scalar linear differential error correcting codes correspond to representable matroid with certain properties. This sheds lots of light in construction and existence of differential error correcting codes for an index coding problem. The definition of a matroidal differential error correcting index code is more general and the matroid need not be a representable matroid. Using a non representable matroid satisfying the conditions, the possibility of non linear differential error correcting codes could be explored.

\appendices
\section{Proof of Lemma \ref{lem:errorCorrection}}
\label{App:LemmaProof}
\textit{Lemma 1 :}
Let $\mathcal{I}_{f(i)}$ denote $(|\overline{\chi_{i}}|+2\delta_{i})$ length vector with a $1$ in one of the first $|\overline{\chi_{i}}|$ position corresponding to the demand $f(i)$ of the receiver $R_{i}$, and with all other elements $0$. For some $ supp(e) \in \lbrace \mathcal{F}\subseteq \lceil N \rfloor : |\mathcal{F}|=2\delta_{i} \rbrace$ the condition 
\begin{equation}
\begin{split}
(z\quad \overline{e})\left(\begin{array}{c}
L\\
\mathbb{I}_{supp(e)}
\end{array}\right)\neq 0, \forall ~ z \in \mathbb{F}_{q}^{n}: z_{\chi_{i}} = 0,z_{f(i)} \neq 0,\\ \forall ~ \overline{e} \in \mathbb{F}_{q}^{2\delta_{i}}
\end{split}
\end{equation}
holds if and only if the following condition holds 
\begin{equation}
\mathcal{I}_{f(i)}\subseteq \left\langle \left(\begin{array}{c}
L_{\overline{\chi_{i}}}\\
\mathbb{I}_{supp(e)}
\end{array}\right)\right\rangle. 
\end{equation}

\begin{proof}
The \textit{if} part is proved first. Since $\mathcal{I}_{f(i)}$ is in subspace of $\left\langle \left(\begin{array}{c}
L_{\overline{\chi_{i}}}\\
\mathbb{I}_{supp(e)}
\end{array}\right)\right\rangle$, linear combinations of columns of $ \left(\begin{array}{c}
L_{\overline{\chi_{i}}}\\
\mathbb{I}_{supp(e)}
\end{array}\right)$ should generate $\mathcal{I}_{f(i)}$. There should be some $N \times 1$ vector $X$ such that, \[ \left(\begin{array}{c}
L_{\overline{\chi_{i}}}\\
\mathbb{I}_{supp(e)}
\end{array}\right) X= \mathcal{I}_{f(i)}. \] Now suppose for some $(z\quad \overline{e})$, with $z_{f(i)}\neq 0,z_{\chi_{i}}=0$ and some $\overline{e} \in \mathbb{F}_{q}^{2\delta_{i}}$ we have \[
(z\quad \overline{e})\left(\begin{array}{c}
L\\
\mathbb{I}_{supp(e)}
\end{array}\right)= 0. \] Since $z_{\chi_{i}}=0$, the above equation reduces to \[
(z_{\overline{\chi_{i}}}\quad \overline{e})\left(\begin{array}{c}
L_{\overline{\chi_{i}}}\\
\mathbb{I}_{supp(e)}
\end{array}\right)= 0. \] Multiplying both sides by $X$, we get $z_{f(i)}=0$, which is a contradiction. This completes the if part. \\
Now we prove the only if part. Let $L_{f(i)}$ denote the row of  $L$ corresponding to the message demanded by receiver $R_{i}$. Let $\overline{r_{i}}$ denote the set $\lceil n \rfloor - \chi_{i} - f(i)$. Let $L_{\overline{r_{i}}} $ denote the submatrix of $L$ with rows indexed by the set $\overline{r_{i}}$. Because (\ref{eq:EC3}) holds, we have \begin{eqnarray*}
rank\left(\begin{array}{c}
L_{\overline{\chi_{i}}}\\
\mathbb{I}_{supp(e)}
\end{array}\right) & = & rank\left(\begin{array}{c}
L_{f(i)}\\
L_{\overline{r_{i}}}\\
\mathbb{I}_{supp(e)}
\end{array}\right)\\
 & = & rank\left(L_{f(i)}\right)+rank\left(\begin{array}{c}
L_{\overline{r_{i}}}\\
\mathbb{I}_{supp(e)}
\end{array}\right)\\
 & = & 1+rank\left(\begin{array}{c}
L_{\overline{r_{i}}}\\
\mathbb{I}_{supp(e)}
\end{array}\right). 
\end{eqnarray*}
Consider the concatenated matrix $\left(\begin{array}{cc}
L_{\overline{\chi_{i}}}\\
 & I_{f(i)}\\
\mathbb{I}_{supp(e)}
\end{array}\right)$, denoted by $Y$. We have,
\begin{eqnarray*}
rank\left(Y\right) & = & rank\left(\begin{array}{cc}
L_{f(i)} & 1\end{array}\right)+rank\left(\begin{array}{c}
L_{\overline{r_{i}}}\\
\mathbb{I}_{supp(e)}
\end{array}\right)\\
 & = & 1+rank\left(\begin{array}{c}
L_{\overline{r_{i}}}\\
\mathbb{I}_{supp(e)}
\end{array}\right).
\end{eqnarray*}
 The concatenated matrix $Y$ has the same rank as the matrix  $\left(\begin{array}{c}
L_{\overline{\chi_{i}}}\\
\mathbb{I}_{supp(e)}
\end{array}\right)$. This proves the only if part. 

\end{proof}

\section{Proof of Theorem \ref{thm:MatroidalErrorCorrecting}}
\label{App:TheoremProof}

\textit{Theorem 1:}
Let $(m,n,\chi,f)$ be an index coding problem. A scalar linear differential $\delta_{R}$ error correcting index code over $\mathbb{F}_{q}$ exists if and only if the index code is matroidal differential $\delta_{R}$ error correcting associated with a $\mathbb{F}_{q}$ representable matroid.
\begin{proof}
First we prove the only if part. Suppose there exists a scalar linear differential error correcting index code $\mathfrak{C}$ of length $N$ over $\mathbb{F}_{q}$ for the index coding problem $(m,n,\chi,f)$. Since the index code is linear it is represented by a $n \times N$ matrix $L$. Let $\zeta$ be the concatenated matrix $\left(\begin{array}{c}
L\\
I_{N}
\end{array}\right)$. Note that order of $\zeta$ is $(n+N) \times N$. Let $\mathcal{Y}$ be the column wise concatenated matrix $\left(\begin{array}{cc}
I_{n+N} & \zeta \end{array}\right)$ of size $(n+N)\times (n+2N)$.  Let $\mathcal{M}=\mathcal{M}(\mathcal{Y})$, the vector matroid associated with $\mathcal{Y}$, with $E(\mathcal{M})$ being the set of column indices of $\mathcal{Y}$. Clearly rank of matroid $\mathcal{M}$ is $n+N$. The function $g: \lceil n \rfloor \cup S(\mathfrak{C}) \rightarrow E(\mathcal{M})$ is defined as follows,\[ g(i)=i, \forall ~i \in \lceil n \rfloor \]
\[ g(c_{i})=n+N+i, \forall ~i \in \lceil N\rfloor. \]
We have to show that the matroid $\mathcal{M}$ and function $g$ satisfies the conditions of Definition \ref{def:MECIC}. Clearly $g(i)$ is one-one for $i \in \lceil n \rfloor$ and $g(\lceil n \rfloor)$ corresponds to the first $n$ columns of $\mathcal{Y}$ which are columns of identity matrix. So $g(\lceil n \rfloor) \in \mathcal{I}(\mathcal{M})$ which proves Condition (A). \\
Consider the basis of $\mathcal{M}$, $B=\lbrace 1,2,\ldots,n+N\rbrace$ obtained by extending $g(\lceil n \rfloor)$. From the definition of $g$, we have $g(c_{i})=n+N+i$ and the corresponding vector associated with it is $\mathcal{Y}^{(n+N+i)}$. From the definition of $\mathcal{Y}$, we have $\mathcal{Y}^{(n+N+i)}=\zeta^{(i)}$. Note that $\zeta^{(i)}=\left(\begin{array}{c}
L^{(i)}\\
I_{N}^{(i)}
\end{array}\right)$, and $\zeta^{(i)}$ does not belong to the linear span of $\mathcal{Y}^{B -  g(\lceil n \rfloor)}$ because $L^{(i)}$ is non zero. Column $L^{(i)}$ corresponds to the $i^{th}$ transmission of the index code and for a scalar linear code it is a non-trivial linear combination of source messages. So condition (B1) holds. The vector $\zeta^{(i)}$ lies in the linear span of $\left(\begin{array}{cc}
I^{\lceil n \rfloor}_{n+N} & I^{n+i}_{(n+N)}\end{array}\right)$ because $L^{(i)}$ lies in the linear span of $I_{n+N}^{\lceil n \rfloor}$ and the vector $ I^{(n+i)}_{n+N}$ has a $1$ at $(n+i)^{th}$ position. So condition (B2) holds.\\
Consider a receiver $R_{i}$ and an error pattern $\mathcal{F}_{j}=\lbrace e_{i_{1}},e_{i_{2}},\ldots,e_{i_{2\delta_{i}}}\rbrace$. Let $I(\mathcal{F}_{j})=\lbrace i_{1},i_{2},\ldots i_{2\delta_{i}} \rbrace$ be the set of indices corresponding to the error pattern and let the set $\lbrace n+i_{1},n+i_{2},\ldots n+i_{2\delta_{i}} \rbrace$ be denoted as $n+I(\mathcal{F}_{j})$. From the definition of $\mathcal{M}_{\mathcal{F}_{j},i}$ we note that it is a vector matroid of the matrix \[
\mathcal{Z}=\mathcal{Y}_{g(\overline{\chi_{i}}) \cup (n+ I(\mathcal{F}_{j}))}.
\]
Note that from the definition of $g$, the matrix that corresponds to $g(S(\mathfrak{C}))$ is \[\mathcal{Z}^{g(S(\mathfrak{C}))} = \zeta_{g(\overline{\chi_{i}}) \cup (n+ I(\mathcal{F}))} = \left(\begin{array}{c}
L_{\overline{\chi_{i}}}\\
I_{I(\mathcal{F}_{j})}
\end{array}\right). \]
Since we have a scalar linear differential error correcting index code, from Lemma \ref{lem:errorCorrection}, we have \[
\mathcal{I}_{f(i)}\subseteq \left\langle \left(\begin{array}{c}
L_{\overline{\chi_{i}}}\\
I_{supp(e)}
\end{array}\right)\right\rangle. 
\]
Note that $f(i) \in \lceil n \rfloor$ and therefore $g(f(i))=f(i)$. So we have $\mathcal{Z}^{(g(f(i)))}=\mathcal{Z}^{(f(i))}$ which is equal to $\mathcal{I}_{f(i)}$. Thus $\mathcal{Z}^{(g(f(i)))}$ lies in the linear span of $\mathcal{Z}^{g(S(\mathfrak{C}))}$ and Condition (C) holds for error pattern $\mathcal{F}_{j}$ and receiver $R_{i}$. Since the receiver and error pattern was chosen arbitrarily this completes the only if part of the proof.

Now we have to prove the if part. Let $\mathcal{M}$ be the given $\mathbb{F}_{q}$ representable matroid with function $g$ and basis $B=g(\lceil n \rfloor) \cup \lbrace b_{n+1},b_{n+2},\ldots b_{n+N} \rbrace$, that satisfy the set of conditions. Let $\mathcal{Y}=\left(\begin{array}{cc}
I_{n+N} & \zeta \end{array}\right)$ be a representation of $\mathcal{M}$ over $\mathbb{F}_{q}$ such that $B=\lbrace 1,2,\ldots,n+N\rbrace$. First we prove that there exists an $n \times N$ matrix $L$ such that $\zeta = 
\left(\begin{array}{c}
L\\
I_{N}
\end{array}\right)$. 
\\Consider a symbol in the index code $c_{j} \in \mathfrak{C}$. From condition (B2) we have that the column vector representing $g(c_{j})$ lies in the linear span of vectors representing $g(\lceil n \rfloor)$ and $b_{n+i}$. This also ensures that $g(c_{j})\notin B$. The vector representing $g(c_{j})$ is $\mathcal{Y}^{(g(c_{j}))}$. So we have \[
\mathcal{Y}^{(g(c_{j}))}= \underset{i \in \lceil n \rfloor}{\sum}a_{i,j} \mathcal{Y}^{(g(i))} + d_{j}\mathcal{Y}^{(n+j)},
\] for some $a_{i,j}$ and $d_{j}$ in $\mathbb{F}_{q}$. As Condition (B1) holds, at least one of the $a_{i,j} \neq 0, \forall ~i \in \lceil n \rfloor$. Condition (B2) also ensures that $d_{j} \neq 0, \forall ~j \in \lceil N \rfloor$. This also ensures that $g(c_{i}) \neq g(c_{j})$ for distinct $c_{i},c_{j} \in \mathfrak{C}$. Arranging all $\mathcal{Y}^{(g(c_{j}))}$, we get $\mathcal{Y}^{g(\mathfrak{C})}=\zeta$, and \[
\zeta = \left(\begin{array}{c}
L_{n\times N}\\
K_{N \times N}
\end{array}\right),
\] where $L$ comprises of the elements $a_{i,j},1\leq i \leq n,1 \leq j \leq N$ and $K$ is a diagonal matrix with $d_{j} , 1\leq j \leq N$ as its diagonal entries. The matroid $\mathcal{M}$ does not change if some row or some column of its representation is multiplied by a non-zero element of $\mathbb{F}_{q}$. The matrix $\mathcal{Y}$ is now of the form $\mathcal{Y}=\left(\begin{array}{cc}
I_{n+N} & \zeta \end{array}\right)$. Consider the matrix $\mathcal{Y}'$ obtained from $\mathcal{Y}$ by multiplying the rows $\lbrace n+1,n+2,\ldots,n+N \rbrace$ by the elements $\lbrace d_{1}^{-1},d_{2}^{-1},\ldots,d_{N}^{-1} \rbrace$ respectively and then multiplying columns $\lbrace n+1,n+2,\ldots,n+N \rbrace$ by $\lbrace d_{1},d_{2},\ldots,d_{N} \rbrace$ respectively.The matrix $\mathcal{Y}'$ is of the form $\left(\begin{array}{cc}

I_{n+N} & \zeta' \end{array}\right)$ where $\zeta' = \left(\begin{array}{c}
L_{n\times N}\\
I_{N}
\end{array}\right)$. The matrix $\mathcal{Y}'$ is a representation for the matroid $\mathcal{M}$ proving our claim. In the last part of the proof we show that the matrix $L$ corresponds to a differential error correcting index code. 
\\
Consider a receiver $R_{i}$ and an arbitrary error pattern $\mathcal{F}_{j}$. For this the matroid $\mathcal{M}_{\mathcal{F}_{j},i}$ is a vector matroid of the matrix \[
\mathcal{Z}=\mathcal{Y}'_{g(\overline{\chi_{i}}) \cup (n+I(\mathcal{F}_{j}))}= \left(\begin{array}{cc}

I_{g(\overline{\chi_{i}}) \cup (n+I(\mathcal{F}_{j}))} & \zeta'_{g(\overline{\chi_{i}}) \cup (n+I(\mathcal{F}_{j}))} \end{array}\right),
\] where ${I}(\mathcal{F}_{j})$ is the index set corresponding to error pattern $\mathcal{F}_{j}$ and $\zeta'_{g(\overline{\chi_{i}}) \cup (n+I(\mathcal{F}_{j}))}= \left(\begin{array}{c}
L_{g(\overline{\chi_{i}})}\\
I_{I(\mathcal{F}_{j})}
\end{array}\right)$. From Condition (C), we have $\mathcal{Z}^{(g(f(i)))} \subseteq \langle \mathcal{Z}^{g(S(\mathfrak{C}))}\rangle$. Note that $\mathcal{Z}^{g(S(\mathfrak{C}))}= \zeta'_{g(\overline{\chi_{i}})\cup (n+I(\mathcal{F}_{j}))}$ and that $\mathcal{Z}^{(g(f(i)))}=\mathcal{I}_{f(i)}$. Thus $\mathcal{I}_{f(i)} \subseteq \left(\begin{array}{c}
L_{g(\overline{\chi_{i}})}\\
I_{I(\mathcal{F}_{j})}
\end{array}\right)$. As the choice of receiver and the error pattern was arbitrary, using Lemma 1 it is seen that the index code given by the matrix $L$ is differential error correcting. This completes the proof of the theorem.
\end{proof}

\section{Example of an error correcting Index Code}
\label{App:ExampleErrorCorrectingIndexCodes}
\begin{example}
\label{eg:example1}
Let $q=2, m=n=3$ and $f(i)=i$ for $i \in \lceil 3 \rfloor $. Let $\chi_{1}=\lbrace 2,3 \rbrace, \chi_{2}=\lbrace 1,3 \rbrace$ and $\chi_{3}=\lbrace 1,2 \rbrace$. Consider the scalar linear index code represented by the matrix \[
L=\left[\begin{array}{cccc}
1 & 1 & 1  \\
1 & 1 & 1  \\
1 & 1 & 1  
\end{array}\right]
.\] Let $\mathfrak{C}$ be the index code based on $L$. We have $\mathfrak{C}(x)= \lbrace x_{1}+x_{2}+x_{3},x_{1}+x_{2}+x_{3},x_{1}+x_{2}+x_{3} \rbrace$ and $S(\mathfrak{C})=\lbrace c_{1},c_{2},c_{3} \rbrace $. It was shown in \cite{ECIC} that the index code corresponding to $L$ is capable of correcting single error. We show that the index code is a matroidal differential $\delta_{R}$ error correcting index code associated with a $\mathbb{F}_{2}$ representable matroid, where $\delta_{R}=\lbrace 1,1,1 \rbrace$. In order to show that we first specify the representable matroid and show the function mapping messages and transmitted symbols to the ground set of matroid. Consider the vector matroid $\mathcal{M}$ associated with the $6 \times 9$ matrix \[
A=\left[\begin{array}{ccccccccc}
1 & 0 & 0 & 0 & 0 & 0 & 1 & 1 & 1  \\
0 & 1 & 0 & 0 & 0 & 0 & 1 & 1 & 1  \\
0 & 0 & 1 & 0 & 0 & 0 & 1 & 1 & 1  \\
0 & 0 & 0 & 1 & 0 & 0 & 1 & 0 & 0  \\
0 & 0 & 0 & 0 & 1 & 0 & 0 & 1 & 0  \\
0 & 0 & 0 & 0 & 0 & 1 & 0 & 0 & 1  \\
\end{array}\right]
.\] The edge set of matroid, $E(\mathcal{M})$ is $\lceil 9 \rfloor$. The function $g: \lceil 3 \rfloor \cup S(\mathfrak{C}) \rightarrow E(\mathcal{M})$, is defined as follows,  \[ g(i)=i, \forall ~ i \in \lceil 3 \rfloor \]
\[ g(c_{i})=6+i, \forall ~ i \in \lceil 3 \rfloor. \] We show that matroid $\mathcal{M}$ and function $g$ satisfies conditions of Definition \ref{def:MECIC}. The first three columns corresponding to the messages are linearly independent. This satisfies Condition A. Let us consider the basis $B=\lceil 6 \rfloor$. The set $B - g(\lceil 3 \rfloor)=\lbrace 4,5,6 \rbrace$. The vector corresponding to $g(c_{i})$ does not lie in the column space of the matrix $A^{B - g(\lceil 3 \rfloor)}$. Thus Condition (B1) is satisfied. Condition (B2) is actually three conditions corresponding to each $c_{i}, i \in \lceil 3 \rfloor$. For $c_{1}$, we have, \[
A^{(g(\lceil 3 \rfloor) \cup b_{4} \cup g(c_{1}))}= \left[\begin{array}{ccccccccc}
1 & 0 & 0 & 0 & 1 \\
0 & 1 & 0 & 0 & 1 \\
0 & 0 & 1 & 0 & 1 \\
0 & 0 & 0 & 1 & 1 \\
0 & 0 & 0 & 0 & 0 \\
0 & 0 & 0 & 0 & 0 \\
\end{array}\right]
.
\] It is clear that the fifth column of $A^{(g(\lceil 3 \rfloor) \cup b_{4} \cup g(c_{1}))}$ is a linear combination of the first four columns and also the fourth column is a linear combination of the remaining four columns. Similarly condition (B2) can be verified for $c_{2}$ and $c_{3}$. 
Condition (C) needs to be verified for various error patterns. Let $\mathcal{F}_{i,j}$ represent the $j^{th}$ error patter at receiver $R_{i}$. For example the error pattern $\mathcal{F}_{1,1}=\lbrace e_{1},e_{2} \rbrace$ is the first error pattern at receiver $R_{1}$. Note that $\delta_i=1$ for all receivers as we are using a single error correcting code. For receiver $R_{1}$, $\overline{\chi_{1}}=\lbrace 1 \rbrace $. We have,
\begin{flalign*}
B_{\overline{\mathcal{F}_{1,1}},1}  & = B - g(\overline{\chi_{1}}) - \lbrace 4,5 \rbrace  \\
& = \lbrace 2,3,6 \rbrace.
\end{flalign*}
The matroid $\mathcal{M}_{\mathcal{F}_{1,1},1}=\mathcal{M}/B_{\overline{\mathcal{F}_{1,1}},1}$, is a vector matroid. Let $M_{i,j}$ be the matrix used for representation of the matroid $\mathcal{M}_{\mathcal{F}_{i,j},i} $. The vector representation for this matroid is given by the matrix \[
M_{1,1}=\left[\begin{array}{ccccccccc}
1 & 0 & 0 & 1 & 1 & 1\\
0 & 1 & 0 & 1 & 0 & 0\\
0 & 0 & 1 & 0 & 1 & 0
\end{array}\right]
.\]
In order to satisfy Condition (C), we want the columns of $M_{1,1}$ corresponding to $g(f(1))$ to be in the linear span of columns of $M_{1,1}$ corresponding to $g(S(\mathfrak{C}))$. The corresponding columns are given below: \[
M_{1,1}^{g(f(1))}=\left[\begin{array}{c}
1 \\
0 \\
0
\end{array}
\right],M_{1,1}^{g(S(\mathfrak{C}))}=\left[\begin{array}{ccc}
1 & 1 & 1 \\
1 & 0 & 0 \\
0 & 1 & 0
\end{array}\right]
.\]
It is clear that columns corresponding to $g(f(1))$ lies in the linear span of columns corresponding to $g(S(\mathfrak{C}))$. Condition (C) is verified for receiver $R_{1}$ and error pattern $\lbrace e_1,e_2 \rbrace$. Let us consider another error pattern $\mathcal{F}_{1,2}=\lbrace e_{1},e_{3} \rbrace$. The matroid $\mathcal{M}_{\mathcal{F}_{1,2},1}=\mathcal{M}/B_{\overline{\mathcal{F}_{1,2}},1}$ is the matroid which we have to consider for error pattern $\mathcal{F}_{1,2}$ and at Receiver $R_{1}$. The vector representation of the matroid $\mathcal{M}_{\mathcal{F}_{1,2},1}$  is given by the matrix \[
M_{1,2}=\left[\begin{array}{ccccccccc}
1 & 0 & 0 & 1 & 1 & 1\\
0 & 1 & 0 & 1 & 0 & 0\\
0 & 0 & 1 & 0 & 0 & 1
\end{array}\right]
.\] We can verify that the first column of matrix $M_{1,2}$ can be obtained as the linear combination of last three columns of matrix. The last error patter to consider is $\mathcal{F}_{1,3}= \lbrace e_{2},e_{3} \rbrace$. Here we consider the matroid $\mathcal{M}_{\mathcal{F}_{1,3},1}$, whose representation is given by the matrix \[
M_{1,3}=\left[\begin{array}{ccccccccc}
1 & 0 & 0 & 1 & 1 & 1\\
0 & 1 & 0 & 0 & 1 & 0\\
0 & 0 & 1 & 0 & 0 & 1
\end{array}\right]
.\] The column corresponding to $g(f(1))$ is the first column and the columns corresponding to $g(S(\mathfrak{C}))$ are last three columns. The linear dependence is very clear in this scenario also. Condition (C) can be verified for other receivers. The matrices corresponding to various error patterns at other receivers are given in Table \ref{Tab:MatricesofExample1}. Condition (C) at all receivers can be verified using Table \ref{Tab:MatricesofExample1}.

\end{example}

\begin{table}
\scriptsize
\begin{tabular}{|c|c|c|c|}
\hline 
\multirow{2}{*}{$(i,j)$} & \multirow{2}{*}{$M_{i,j}$} & \multirow{2}{*}{$M_{i,j}^{f(i)}$} & \multirow{2}{*}{$M_{i,j}^{g(\mathfrak{C}(x))}$} \\
& &  &  \\
\hline
& & &\\
$(1,1)$ & 
$\left[\begin{array}{cccccc}
1 & 0 & 0 & 1 & 1 & 1\\
0 & 1 & 0 & 1 & 0 & 0\\
0 & 0 & 1 & 0 & 1 & 0
\end{array}\right]$ & $\left[\begin{array}{c}
1\\
0\\
0
\end{array}\right]$ & $\left[\begin{array}{ccc}
1 & 1 & 1\\
1 & 0 & 0\\
0 & 1 & 0
\end{array}\right]$ \\
& & & \\
\hline 
& & & \\
$(1,2)$ &
$\left[\begin{array}{ccccccccc}
1 & 0 & 0 & 1 & 1 & 1\\
0 & 1 & 0 & 1 & 0 & 0\\
0 & 0 & 1 & 0 & 0 & 1
\end{array}\right]$ & $\left[\begin{array}{c}
1 \\
0 \\
0
\end{array}
\right]$ & $\left[\begin{array}{ccc}
1 & 1 & 1 \\
1 & 0 & 0 \\
0 & 0 & 1
\end{array}\right]$ \\
& & &\\
\hline 
& & & \\
$(1,3)$ &
$\left[\begin{array}{ccccccccc}
1 & 0 & 0 & 1 & 1 & 1\\
0 & 1 & 0 & 0 & 1 & 0\\
0 & 0 & 1 & 0 & 0 & 1
\end{array}\right]$ & $\left[\begin{array}{c}
1 \\
0 \\
0
\end{array}
\right]$ & $\left[\begin{array}{ccc}
1 & 1 & 1 \\
0 & 1 & 0 \\
0 & 0 & 1
\end{array}\right]$ \\
& & &\\
\hline 
& & & \\
$(2,1)$ &
$\left[\begin{array}{ccccccccc}
1 & 0 & 0 & 1 & 1 & 1\\
0 & 1 & 0 & 1 & 0 & 0\\
0 & 0 & 1 & 0 & 1 & 0
\end{array}\right]$ & $\left[\begin{array}{c}
1 \\
0 \\
0
\end{array}
\right]$ & $\left[\begin{array}{ccc}
1 & 1 & 1 \\
1 & 0 & 0 \\
0 & 1 & 0
\end{array}\right]$ \\
& & &\\
\hline 
& & & \\
$(2,2)$ &
$\left[\begin{array}{ccccccccc}
1 & 0 & 0 & 1 & 1 & 1\\
0 & 1 & 0 & 1 & 0 & 0\\
0 & 0 & 1 & 0 & 0 & 1
\end{array}\right]$ & $\left[\begin{array}{c}
1 \\
0 \\
0
\end{array}
\right]$ & $\left[\begin{array}{ccc}
1 & 1 & 1 \\
1 & 0 & 0 \\
0 & 0 & 1
\end{array}\right]$ \\
& & &\\
\hline 
& & & \\
$(2,3)$ &
$\left[\begin{array}{ccccccccc}
1 & 0 & 0 & 1 & 1 & 1\\
0 & 1 & 0 & 0 & 1 & 0\\
0 & 0 & 1 & 0 & 0 & 1
\end{array}\right]$ & $\left[\begin{array}{c}
1 \\
0 \\
0
\end{array}
\right]$ & $\left[\begin{array}{ccc}
1 & 1 & 1 \\
0 & 1 & 0 \\
0 & 0 & 1
\end{array}\right]$ \\
& & &\\
\hline 
& & & \\
$(3,1)$ &
$\left[\begin{array}{ccccccccc}
1 & 0 & 0 & 1 & 1 & 1\\
0 & 1 & 0 & 1 & 0 & 0\\
0 & 0 & 1 & 0 & 1 & 0
\end{array}\right]$ & $\left[\begin{array}{c}
1 \\
0 \\
0
\end{array}
\right]$ & $\left[\begin{array}{ccc}
1 & 1 & 1 \\
1 & 0 & 0 \\
0 & 1 & 0
\end{array}\right]$ \\
& & &\\
\hline 
& & & \\
$(3,2)$ &
$\left[\begin{array}{ccccccccc}
1 & 0 & 0 & 1 & 1 & 1\\
0 & 1 & 0 & 1 & 0 & 0\\
0 & 0 & 1 & 0 & 0 & 1
\end{array}\right]$ & $\left[\begin{array}{c}
1 \\
0 \\
0
\end{array}
\right]$ & $\left[\begin{array}{ccc}
1 & 1 & 1 \\
1 & 0 & 0 \\
0 & 0 & 1
\end{array}\right]$ \\
& & &\\
\hline 
& & & \\
$(3,3)$ &
$\left[\begin{array}{ccccccccc}
1 & 0 & 0 & 1 & 1 & 1\\
0 & 1 & 0 & 0 & 1 & 0\\
0 & 0 & 1 & 0 & 0 & 1
\end{array}\right]$ & $\left[\begin{array}{c}
1 \\
0 \\
0
\end{array}
\right]$ & $\left[\begin{array}{ccc}
1 & 1 & 1 \\
0 & 1 & 0 \\
0 & 0 & 1
\end{array}\right]$ \\
& & &\\
\hline 
\end{tabular}
\caption{\footnotesize Matrices representing the contracted matroids used in Example 1. Matrix $M_{i,j}$ is the representation of the matroid $\mathcal{M}_{\mathcal{F}_{i,j},i} $ }
\label{Tab:MatricesofExample1}
\normalsize
\end{table}

\section{Example of an index code capable of error correcting at a subset of receivers}
\label{App:ExampleErrorCorrectionSubsetofReceivers}
\begin{example}
\label{eg:example2}
In this example we consider the index coding problem of Example \ref{eg:example1}. However in this example only receiver $R_{1}$ requires single error correction. Consider another index code $\mathfrak{C}_{2}$ described by the matrix \[
L_{1}=\left[\begin{array}{cccc}
1 & 1 & 1  \\
1 & 1 & 0  \\
1 & 0 & 1  
\end{array}\right]
.\] We have $\mathfrak{C}_{2}(x)=\lbrace x_{1}+x_{2}+x_{3},x_{1}+x_{2},x_{1}+x_{3} \rbrace$. Consider the representable matroid $\mathcal{M}'$ associated with the $6 \times 9$ matrix\[
A_{1}=\left[\begin{array}{ccccccccc}
1 & 0 & 0 & 0 & 0 & 0 & 1 & 1 & 1  \\
0 & 1 & 0 & 0 & 0 & 0 & 1 & 1 & 0  \\
0 & 0 & 1 & 0 & 0 & 0 & 1 & 0 & 1  \\
0 & 0 & 0 & 1 & 0 & 0 & 1 & 0 & 0  \\
0 & 0 & 0 & 0 & 1 & 0 & 0 & 1 & 0  \\
0 & 0 & 0 & 0 & 0 & 1 & 0 & 0 & 1  \\
\end{array}\right]
.\] The function $g: \lceil 3 \rfloor \cup S(\mathfrak{C}_{2}) \rightarrow E(\mathcal{M})$, is defined as follows,  \[ g(i)=i, \forall ~ i \in \lceil 3 \rfloor \]
\[ g(c_{i})=6+i, \forall ~ i \in \lceil 3 \rfloor. \] Condition A and condition B of Definition \ref{def:MECIC} holds true for this matroid. Let us consider receiver $R_{1}$, basis $B= \lceil 6 \rfloor $ and error pattern $\mathcal{F}_{1,1}=\lbrace e_{1}, e_{2} \rbrace$. For receiver $R_{1}$, $\overline{\chi_{1}}=\lbrace 1 \rbrace .$ We have,
\begin{flalign*}
B_{\overline{\mathcal{F}_{1,1}},1}  & = B - g(\overline{\chi_{1}})- \lbrace 4,5 \rbrace  \\
& = \lbrace 2,3,6 \rbrace.
\end{flalign*} The matroid $\mathcal{M}'_{\mathcal{F}_{1,1},1}=\mathcal{M}'/B_{\overline{\mathcal{F}_{1,1}},1}$ is a vector matroid. Let $M_{i,j}$ be the representation of the matroid $\mathcal{M}'_{\mathcal{F}_{i,j},i}$. The vector representation of the matroid $\mathcal{M}'_{\mathcal{F}_{1,1},1}$ is given by the matrix \[
M_{1,1}=\left[\begin{array}{ccccccccc}
1 & 0 & 0 & 1 & 1 & 1\\
0 & 1 & 0 & 1 & 0 & 0\\
0 & 0 & 1 & 0 & 1 & 0
\end{array}\right]
.\] We can clearly see that Condition (C) is satisfied. The matrices representing the contracted matroids for receiver $R_{1}$ is given in Table \ref{Tab:matricesexample2}. Let us consider Receiver $R_{2}$, basis $B= \lceil 6 \rfloor $ and error pattern $\mathcal{F}_{2,1}=\lbrace e_{1}, e_{2} \rbrace$. For receiver $R_{2}$, $\overline{\chi_{2}}=\lbrace 2 \rbrace .$ We have,
\begin{flalign*}
B_{\overline{\mathcal{F}_{2,1}},2}  & = B\setminus g(\overline{\chi_{2}}) \setminus \lbrace 4,5 \rbrace  \\
& = \lbrace 1,3,6 \rbrace.
\end{flalign*} The matroid $\mathcal{M}'_{\mathcal{F}_{2,1},2}=\mathcal{M}'/B_{\overline{\mathcal{F}_{2,1}},2}$ is a vector matroid. The vector representation of the matroid $\mathcal{M}'_{\mathcal{F}_{2,1},1}$ is given by the matrix \[
M_{2,1}=\left[\begin{array}{ccccccccc}
1 & 0 & 0 & 1 & 1 & 0\\
0 & 1 & 0 & 1 & 0 & 0\\
0 & 0 & 1 & 0 & 1 & 0
\end{array}\right]
.\] In order to satisfy condition (C) we want the columns of $M_{2,1}$ corresponding to $g(f(2))$ to be in the linear span of columns of $M_{2,1}$ corresponding to $g(S(\mathfrak{C}_{2}))$. Observe that $M_{2,1}^{g(f(2))}=\left[\begin{array}{c}
1 \\
0 \\
0
\end{array}
\right]$ and $M_{2,1}^{g(S(\mathfrak{C}_{2}))}=\left[\begin{array}{ccc}
1 & 1 & 0 \\
1 & 0 & 0 \\
0 & 1 & 0
\end{array}\right]$. The rank of the columns corresponding to $g(S(\mathfrak{C}_{2}))$ is $2$ where as the rank of the columns corresponding to $g(S(\mathfrak{C}_{2}))$ along with $g(f(2))$ is $3$. Thus condition (C) is violated. So the index code corresponding to matrix $L_{1}$ does not give error correcting property at receiver $R_{2}$. Similarly for receiver $R_{3}$ and error pattern $\mathcal{F}_{3,2}=\lbrace e_{1},e_{3} \rbrace$, condition (C) gets violated. The matrix $M_{3,2}$ representing the matroid $\mathcal{M}'_{\mathcal{F}_{3,2},3}$ is \[
M_{3,2}=\left[\begin{array}{ccccccccc}
1 & 0 & 0 & 1 & 0 & 1\\
0 & 1 & 0 & 1 & 0 & 0\\
0 & 0 & 1 & 0 & 0 & 1
\end{array}\right]
.\] We can observe that condition (C) is not satisfied by receiver $R_{3}$. Thus the index code described by $L_{1}$ has single error correcting property only at receiver $R_{1}$.

\begin{table}
\scriptsize
\begin{tabular}{|c|c|c|c|}
\hline 
\multirow{2}{*}{$(i,j)$} & \multirow{2}{*}{$M_{i,j}$} & \multirow{2}{*}{$M_{i,j}^{f(i)}$} & \multirow{2}{*}{$M_{i,j}^{g(\mathfrak{C}(x))}$} \\
& &  &  \\
\hline
& & &\\
$(1,1)$ & 
$\left[\begin{array}{cccccc}
1 & 0 & 0 & 1 & 1 & 1\\
0 & 1 & 0 & 1 & 0 & 0\\
0 & 0 & 1 & 0 & 1 & 0
\end{array}\right]$ & $\left[\begin{array}{c}
1\\
0\\
0
\end{array}\right]$ & $\left[\begin{array}{ccc}
1 & 1 & 1\\
1 & 0 & 0\\
0 & 1 & 0
\end{array}\right]$ \\
& & & \\
\hline 
& & & \\
$(1,2)$ &
$\left[\begin{array}{ccccccccc}
1 & 0 & 0 & 1 & 1 & 1\\
0 & 1 & 0 & 1 & 0 & 0\\
0 & 0 & 1 & 0 & 0 & 1
\end{array}\right]$ & $\left[\begin{array}{c}
1 \\
0 \\
0
\end{array}
\right]$ & $\left[\begin{array}{ccc}
1 & 1 & 1 \\
1 & 0 & 0 \\
0 & 0 & 1
\end{array}\right]$ \\
& & &\\
\hline 
& & & \\
$(1,3)$ &
$\left[\begin{array}{ccccccccc}
1 & 0 & 0 & 1 & 1 & 1\\
0 & 1 & 0 & 0 & 1 & 0\\
0 & 0 & 1 & 0 & 0 & 1
\end{array}\right]$ & $\left[\begin{array}{c}
1 \\
0 \\
0
\end{array}
\right]$ & $\left[\begin{array}{ccc}
1 & 1 & 1 \\
0 & 1 & 0 \\
0 & 0 & 1
\end{array}\right]$ \\
& & &\\
\hline
\end{tabular}
\\
\caption{\footnotesize Matrices representing the contracted matroids in Example \ref{eg:example2} for receiver $R_{1}$. Matrix $M_{i,j}$ is the representation of the matroid $\mathcal{M}'_{\mathcal{F}_{i,j},i} $ }
\label{Tab:matricesexample2}
\normalsize
\end{table}

\end{example}

\end{document}